\documentclass[12pt]{article}
\usepackage{amstext,amsmath,amsfonts}
\input epsf

\title{Virtual Time Horizon Control via Communication Network 
Design\footnote{Invited review paper for Volume on
Computational Complexity and Statistical Physics,
Santa Fe Institute Studies in the Sciences of Complexity series,
Oxford University Press, 2003.}}
\date{\today}

\author{Zoltan Toroczkai\thanks{Complex Systems Group, Theoretical Division,
Los Alamos National Laboratory, MS-B213, Los Alamos, NM,
87545, USA. {\tt toro@lanl.gov}}
\and
Gy\"orgy Korniss\thanks{Department of Physics, Applied Physics, and
Astronomy, Rensselaer Polytechnic Institute, 110 8th Street, Troy,
NY 12180-3590 USA. {\tt korniss@rpi.edu}}
\and
Mark A. Novotny\thanks{Department of Physics and Astronomy and ERC, 
Mississippi State University, Mississippi State, MS 39762-5167 USA, 
{\tt novotny@erc.msstate.edu}}
\and
Hasan Guclu\thanks{Department of Physics, Applied Physics, and
Astronomy, Rensselaer Polytechnic Institute, 110 8th Street, Troy,
NY 12180-3590 USA. {\tt gucluh@rpi.edu}}}

\begin{document}
\thispagestyle{empty}
\maketitle
\renewcommand{\thefootnote}{}

\renewcommand{\thefootnote}{\arabic{footnote}}

\begin{abstract}
We consider massively parallel discrete event simulations where
the communication topology among the processing elements is a
complex graph. In the case of regular topologies we review recent
results on virtual time horizon management. First we analyze the
computational scalability of the conservative massively parallel
update scheme for discrete event simulations by using the analogy
with a well-known surface growth model, then we show that a simple
modification of the regular PE communication topology to a
small-world topology will also ensure measurement scalability.
This leads to a fully scalable parallel simulation for systems
with asynchronous dynamics and short-range interactions. Finally,
we present numerical results for the evolution of the virtual time
horizon on scale-free Barab\'asi-Albert networks serving as 
communication topology among the processing elements.
\end{abstract}

\newpage

\section{Introduction to the scalability problem of massively
parallel discrete-event simulations}

The description and understanding of complex systems dynamics is in
most cases impossible
via analytic methods. The density of problems that are rigorously
solvable with analytic tools is vanishingly small in the set of all problems. The only
way one can obtain system level understanding of such problems is through
direct simulation.
The class of complex systems considered here are made of a large number of interacting
individual elements with a finite number of attributes, or local state
variables, each assuming
a countable number (typically finite) of values. The dynamics of the local
state variables
are discrete events occurring in continuous time. Between two consecutive
updates, the local variables
stay unchanged. Another important assumption we make is that the interactions
in the underlying system to be simulated have finite range.
Examples of such systems include:
magnetic systems (spin states and spin flip dynamics), surface growth
via molecular beam epitaxy (height of the surface at a given point, measured
from a growth level,
molecular deposition and diffusion dynamics); epidemiology (health of an
individual, health state change
due to infection, or recovery); financial markets (wealth state, buy/sell
dynamics), wireless communications,
or queueing systems (number of jobs, job arrival dynamics).

Often, as the case studied here, the dynamics of such systems is
inherently stochastic and {\em asynchronous}. The simulation of
such systems is rather non-trivial and in most cases the
complexity of the problem requires simulations on distributed
architectures, defining the field of Parallel Discrete-Event
Simulations (PDES) \cite{FUJI90,NICOL94,LUBA00}.  Conceptually,
the computational task is divided among $N$ processing elements
(PE-s), where each processor evolves the dynamics of the allocated
piece. Due to the interactions among the individual elements of
the simulated system (spins, atoms, packets, calls, etc.) the PE-s
must coordinate with a subset of other PE-s during the simulation.
For example, the state of a spin can only be updated if the
state of the neighbors is known. However, some neighbors might
belong to the computational domain of another PE, thus message
passing will be required in order to preserve causality. In the
PDES schemes we analyze, update attempts are self-initiated \cite{FELD91} and
are independent of the configuration of the underlying system
\cite{LUBA87,LUBA88}. Although these properties simplify the
analysis of the corresponding PDES schemes, they can be highly
efficient \cite{KNR99} and are readily applicable to a large number
problems in science and engineering. Further, the performance and
the scalability of these PDES schemes become independent of the
specific underlying system i.e., we learn the generic behavior of
these complex computational schemes.

The update dynamics, together with the information sharing among PE-s, make
the parallel discrete-event
simulation process a complex dynamical system itself. In fact,
it perfectly fits the type
of complex systems we are considering here: the individual elements are the
PE-s, and their states (local simulated time) evolve
according to update events which are dependent on the states of the neighboring PE-s.

With the number and size of parallel computers on the rise,  the problem of
designing efficient parallel
algorithms, or update schemes becomes increasingly important. In passing, we
can mention a few examples
of large parallel computers:  the 9472-node ASCII Red at Sandia, the 12288-node
QCDSP Teraflop Machine at Brookhaven, the 8192-node IBM ASCII
White with 12.3 Teraflops. IBM is currently building the Blue Gene/L with
200 Teraflops,
with over $10^6$ nodes. As a matter of fact the largest supercomputer ever built is by Nature itself:
the brain, which does an immense parallel computing task to sustain the individual.
In particular the human brain has $10^{11}$ PE-s (neurons) each with an average of $10^4$
synaptic connections, creating a bundle on the order of $10^{15}$ ``wires'' 
jammed into a volume of approximately $1400\;\;\mbox{cm}^3$.

The design of efficient parallel update schemes is a rather
challenging problem, due to the fact that the
dynamics of the simulation scheme itself is a
complex system, whose properties are  hard to deduce
using classical methods of algorithm analysis.
Here we present a less conventional approach to the analysis
of efficiency and scalability for the class of
massively parallel conservative PDE-s schemes, by exactly
mapping the parallel computational process itself
onto a non-equilibrium surface growth model \cite{KTNR00}.
This allows us to translate the questions about
efficiency and scalability into questions
formulated in terms of certain
topological properties of this non-equilibrium surface.
Then, using methods from statistical mechanics,
developed some time ago to study the dynamics of
such surfaces (in a completely different context), we
solve the scalability problem of the computational
PDES-s scheme \cite{KTNR00,KNGTR03}.
Similar connections between computational
schemes and complex systems behavior have recently
been made \cite{SOS01,S99} for rollback-based
PDES-s algorithms \cite{J85} and self-organized criticality \cite{BTW87}.

Since one is interested in the {\em dynamics} of the underlying
complex system, the PDES scheme must simulate the `physical time'
variable of the complex system. When the simulations are done on a
single processor machine, a single (global) time stream is
sufficient to ``label'' or time-stamp the updates of the local
configurations, regardless whether the dynamics of the underlying
system is synchronous or asynchronous. When simulating
asynchronous dynamics on distributed architectures, however, each
PE generates its own physical, or virtual time, which is the
physical time variable of the particular computational domain
handled by that PE. Due to the varying complexity of the
computation at different PE-s, at a given wall-clock instant the
simulated virtual times of the PE-s can differ, a phenomenon
called ``time horizon roughening". We denote the simulated, or
virtual time at PE $i$ measured at wall-clock time $t$, by
$\tau_i(t)$. For non-interacting subsystems the wall-clock time 
$t$ is directly proportional to
the (discrete) number of parallel steps simultaneously performed
on each PE, also called the number of Monte-Carlo steps (MCS)
in dynamic Monte Carlo simulations. Without altering the meaning,
$t$ from now on will be used to denote the number of discrete
steps performed in the parallel simulation. The set of virtual
times $\{\tau_i(t)\}_{i=1}^{N}$ forms the {\em virtual time
horizon} of the PDES-s scheme after $t$ parallel updates.

\begin{figure}[htbp]
\protect\vspace*{-0.1cm} \epsfxsize = 5 in
\centerline{\epsfbox{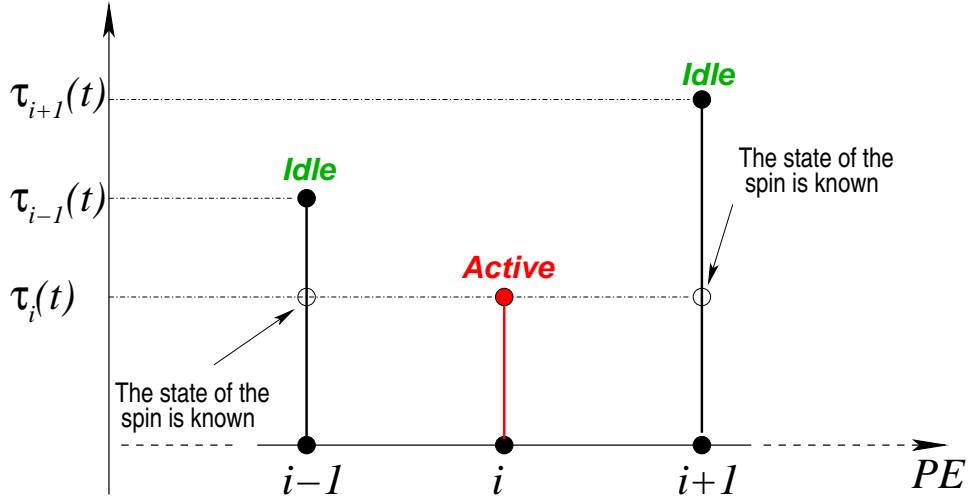}}
\protect\vspace*{-0.2cm}
\caption{A simple diagram to illustrate the conservative PDES-s
scheme for a one-dimensional system with nearest-neighbor interactions.}
\label{Fig:1}
\end{figure}

In conservative PDES-s schemes \cite{CHAND81}, a PE will only
perform its next update if it can obtain the correct information
to evolve the local configuration (local state) of the underlying
physical system it simulates, without violating causality.
Otherwise, it idles. Specifically, when the underlying system has
nearest-neighbor interactions, each PE must check with its
``neighboring" PEs (mimicking the interaction topology of the
underlying system) to see if those are progressed at least up to
the point in virtual time where the PE itself did
\cite{LUBA87,LUBA88}. Based on the fundamental notion of
discrete-event systems that the state of a local state variable
remains unchanged between two successive update attempts, the
above rule guarantees the causality of the simulated dynamics
\cite{LUBA87,LUBA88}. A simple example illustrating this is shown
in Figure \ref{Fig:1}.
One can consider, for example, a magnetic
system as the underlying physical system, where the spins are
arranged in the sites of a one-dimensional lattice, and that a
single spin is handled by a single PE (for more realistic and
efficient implementations see \cite{LUBA87,LUBA88,KNR99}). In Fig.
\ref{Fig:1}, which shows the distribution of the virtual simulated
times at a given wall-clock instant $t$, the only PE that can update
from the set $\{i-1,i,i+1\}$ is in site $i$, since the state of
the neighboring spins at sites $i\pm 1$ are already known.
However, PE-s $i\pm 1$ cannot update their spin states at wall
clock instant $t$, because,  the state of the neighboring spin $i$
at {\em their} simulated times (at $\tau_{i-1}$ and $\tau_{i+1}$)
is not known yet. In other words PE $i$ can only update at
wall-clock instant $t$ if $\tau_i(t) \leq
\min\{\tau_{i-1}(t),\tau_{i-1}(t)\}$, i.e., its virtual time is a
{\em local minimum} among the virtual times of its neighboring
PE-s. It is easy to see that the same conclusion holds for
arbitrary PE topologies. Let the topology for the communication
among the processing elements be symbolized by a graph $G(V,E)$,
where $V$ is the vertex set of $N$ nodes and $E$ is the edge set
of $G$. Given a node $i \in V(G)$, we denote by $S^{(1)}_i$ the
set of first order neighbors on $G$ of $i$. Then, node (PE) $i$
can update its state, in the conservative PDES-s scheme iff:
\begin{equation}
\tau^{(G)}_i(t) \leq \min_{j\in S^{(1)}_i}
\{ \tau_{j}^{(G)}(t) \}\;\;,i=1,..,N. \label{active}
\end{equation}
In the following, the set of (active) nodes which obey condition
(\ref{active}) at time $t$, will be denoted by ${\cal A}(t)$. Now
we are in the position to formulate the scalability problem of
PDES-schemes for systems with asynchronous dynamics. For the
PDES-s scheme to be fully scalable, the following two criteria
must be met: (i) the virtual time horizon must progress on average
at a nonzero rate, and (ii) the typical spread of the time horizon
should be finite, as the number of PE-s $N$ goes to infinity. When
the first criterion is ensured for large enough times $t$, the
simulation is said to be {\em computationally scalable}, and it
just means that when increasing the size of the computation to
infinity, while keeping the average computational domain/load on a
single  PE the same, the simulation will progress at a nonzero
rate. However, as we will show below, increasing the system size,
the {\em spread} in the time horizon can diverge, severely
hindering frequent data collection about the state of the
simulated system. Specifically, when one requires to take a
measurement of some physical property of the simulated system at
(virtual, or simulated) physical time $\tau$, we have to {\em
wait} (in wall-clock time) until all the virtual simulated times
at all the PE-s pass through the value of $\tau$. Thus, in order
to collect system-wide measurements from the simulation, we incur
a waiting time proportional to the {\em spread}, or width of the
fluctuating time horizon. For PDES-s schemes for which the spread
diverges with system size, the waiting time for the measurements
will also diverge, and the scheme is {\em not} measurement
scalable. When condition (ii) is fulfilled for large enough times
$t$, we say that the PDES-s scheme is {\em measurement scalable}.

The  scalability criteria above can simply be formalized in terms of the properties of the virtual time horizon,
$\{\tau^{(G)}_i(t)\}_{i=1}^{N}$. The average of the time horizon after $t$ parallel steps is:
\begin{equation}
\overline{\tau}^{(G)}(t) = \frac{1}{N} \sum_{j=1}^{N} \tau_{j}^{(G)}(t)\;. \label{tavr}
\end{equation}
At a given (wall-clock) time $t$ the only PE-s that
can make progress, i.e., are not in idle, are those which have simulated virtual times obeying condition
(\ref{active}). Thus, the rate of progress of the time horizon average becomes:
\begin{equation}
\overline{\tau}^{(G)}(t+1)-\overline{\tau}^{(G)}(t) = \frac{1}{N} \sum_{l \in {\cal A}(t)}
\left[\tau_{l}^{(G)}(t+1) - \tau_{l}^{(G)}(t)\right]\;. \label{tdiff}
\end{equation}
The difference in the square brackets on the r.h.s of (\ref{tdiff}) is the {\em physical time} elapsed between
two consecutive events in the physical domain simulated by the $l$th PE, and it is determined by the
physical process responsible for the stochastic dynamics of the simulated complex system. If we replace the
time intervals in square brackets in (\ref{tdiff}) with their (obviously finite) average value $\Delta$,
we obtain that the average progress rate of the time horizon, or average {\em utilization}
$\langle u^{(G)}(t) \rangle =
\langle \overline{\tau}^{(G)}(t+1)-\overline{\tau}^{(G)}(t) \rangle$ is proportional to the {\em number}
of non-idling, or active
PE-s. The average $\langle \cdot \rangle$ is taken over the stochastic event dynamics, which is assumed to be
the same at all sites. For many cases, the $\Delta$ factor is independent on $N$
(due to the finite range of the interaction
in the complex system), so the computational efficiency, or average utilization of the simulation can simply be
identified with the average density of the active PE-s:
\begin{equation}
\langle u^{(G)}(t) \rangle = \frac{\langle |{\cal A}^{(G)}(t)| \rangle }{N} \label{u}
\end{equation}
where $|{\cal A}^{(G)}(t)|$ denotes the number of elements of the set ${\cal A}^{(G)}(t)$.  Thus, the PDES-s
scheme is computationally scalable, if there exists a {\em constant} $c > 0$, such that:
\begin{equation}
\langle u^{(G)}(\infty) \rangle =
\lim_{{t \to \infty \atop N\to \infty}}\frac{\langle |{\cal A}^{(G)}(t)| \rangle}{N} > c. \label{uinf}
\end{equation}
The measurement scalability of the PDES-s scheme, is characterized by the spread of the virtual time horizon.
Instead of dealing with the actual spread (difference between the maximum and minimum values) we shall consider
the average ``width'' of the time horizon defined as:
\begin{equation}
\langle [w^{(G)}]^2(t)\rangle = \frac{1}{N} \sum_{j = 1}^{N}
\left[\tau_{j}^{(G)}(t) - \overline{\tau}^{(G)}(t)\right]^2\;. \label{twidth}
\end{equation}

A PDES-s scheme is measurement scalable, if there exists a {\em constant} $M > 0$, such:
\begin{equation}
\langle [w^{(G)}]^2(\infty)\rangle = \lim_{{t \to \infty \atop N\to \infty}}
\frac{1}{N} \sum_{j = 1}^{N}
\left[\tau_{j}^{(G)}(t) - \overline{\tau}^{(G)}(t)\right]^2 < M\;. \label{twidthinf}
\end{equation}

In reality, the number of PEs $N$ or the simulation time $t$ can
never be taken to infinity, so for practical purposes, the
scalability is deduced from the scaling behavior of the quantities
for long times and for large number of PEs. The setup presented
above is perfectly suitable to establish a mapping between
non-equilibrium surface growth models \cite{BS95} and conservative
PDES-s schemes. We discuss extensively this mapping in the next
section.

The paper is organized as follows: in Section \ref{sec2} we
discuss the scalability of the computational phase and the failure
of scalability of the measurement phase of the basic conservative
scheme on regular topologies. Then we show how a simple
modification of the communication topology (from a regular lattice
to a small-world structure) leads to a fully scalable PDES-s
scheme. In section \ref{sec3} we study the scalability problem on
scale-free network topologies. Section \ref{sec4} is devoted to
conclusions.

\section{Scalability of conservative PDES-s schemes on
regular and small-world topologies}\label{sec2}

In many large complex systems the stochastic event dynamics can be
characterized by a Poisson distributed stream. For example, in an
Ising magnet with single spin-flip Glauber dynamics \cite{KNR99}
the spin-flip attempts are Poisson distributed events, or  in
wireless cellular communications the call arrivals also obey
Poisson statistics \cite{GLNW94}, etc. In the following we
restrict ourselves to such Poisson distributed stochastic
processes for event dynamics, however numerical simulations show,
that our conclusions for scalability hold for a large class of
other stochastic distributions, as well. The evolution of the
virtual time horizon incorporating condition
(\ref{active}) for Poisson asynchrony is given by the
equation:
\begin{equation}
\tau_{i}^{(G)}(t+1) = \tau_i^{(G)}(t)+\eta_i(t) \prod_{j\in S^{(1)}_i}
\theta(\tau_j^{(G)}(t)-\tau_i^{(G)}(t)) \label{eveq}
\end{equation}
Here $\theta(x)$ is the Heaviside step function, and
$\eta_i(t)$ the Poisson distributed virtual
time increment at PE $i$, and time $t$ is. These
increments are drawn at random, independently of
$i$ and $t$, and of the existing time horizon.

\subsection{The basic conservative scheme on regular topologies}

Next, we consider the {\em basic} conservative scheme, which is defined on
regular, square lattice communication topologies, in $d$ dimensions, so that $N=L^d$.
For brevity, we drop from the superscript $(G)$ in the notation for $\tau_{i}(t)$. In particular,
we first illustrate our analysis on the simplest regular topology, that of a regular one-dimensional
lattice, with periodic boundary conditions ($G$ is a ring). Later we discuss the
general, $d$-dimensional case. The evolution equation on the ring becomes simply:
\begin{equation}
\tau_{i}(t+1) = \tau_i(t)+\eta_i(t) \theta(\tau_{i-1}(t)-\tau_i(t))
\theta(\tau_{i+1}(t)-\tau_i(t)). \label{ring}
\end{equation}
with the boundary conditions $\tau_{N+1} = \tau_{N} = \tau_0$. The total number of active sites/PE-s
is thus given by
$|{\cal A}(t)| = \sum_{i=1}^{N} \theta(\tau_{i-1}(t)-\tau_i(t))\theta(\tau_{i+1}(t)-\tau_i(t))$
so the average utilization (\ref{u})
becomes:
\begin{equation}
\langle u(L,t) \rangle =  \frac{1}{N} \sum_{i=1}^{N} \left\langle\theta(\tau_{i-1}(t)-\tau_i(t))
\theta(\tau_{i+1}(t)-\tau_i(t))\right\rangle \label{utiR}
\end{equation}
The average $\langle \cdot \rangle$ is performed over the random variables $\{ \eta_i(t') \}_{{i=1,..,L
\atop t' = 1,..,t }}$ which have an exponential distribution, $Prob\{ x < \eta \leq x+\delta x \}  =
\int_{x}^{x+\delta x} dy e^{-y}$. In spite of the simple appearance of the dynamics (\ref{ring}), and
the exponential (or Poisson) stochastic dynamics at nodes, calculating the average utilization
(\ref{utiR}) is very difficult. Even a rigorous proof for the existence of the
lower bound (\ref{uinf}) using direct methods is still an open problem.

\begin{figure}[htbp]
\protect\vspace*{-0.1cm} \epsfxsize = 4 in
\centerline{\epsfbox{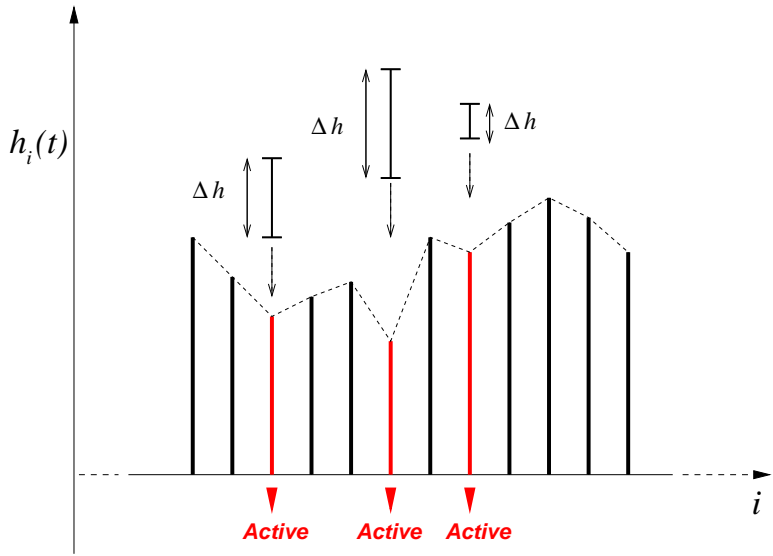}}
\protect\vspace*{-0.2cm}
\caption{A simple surface growth model on a 1-$d$
substrate corresponding to the basic conservative
PDES-s scheme.}
\label{Fig:2}
\end{figure}

Here we present a different approach,
by mapping first the problem to a non-equilibrium surface grown via molecular beam epitaxy
(where atoms or molecules are deposited from vapors, or beams onto the surface) model.
The quantities brought in analogy
are: the $i$-th PE is the site $i$ in the substrate; the number of parallel updates $t$ is the number of
deposited monolayers; $\tau_i(t)$ is the height $h_i(t)$ at site $i$ and time $t$; a virtual time increment
of $\eta_i(t)$ at PE $i$ in the $t$-th step, corresponds to a material ``rod'' of length $\eta_i(t)$ deposited onto
the surface, see Figure (\ref{Fig:2}). The length of the rod is a Poisson distributed random variable.
During the $t$-th update, the rods are deposited only in {\em local minima} of the surface.
The utilization of the PDES-s scheme corresponds to the the {\em density} of local minima of the growing surface.
Even though the length of the rods are random independent variables, the fact that they can only be deposited
in local minima will generate lateral correlations into the surface fluctuations, and makes the problem hard to
compute exactly.
The rods are deposited onto the surface in a parallel update scheme: after all local minima are updated
(deposited onto) the the time $t$ is incremented by unity. We will coin the surface growth analog of our
basic conservative PDES-s scheme , the `MPEU' model (Massively Parallel Exponential Update Model).

\begin{figure}[htbp]
\protect\vspace*{-0.1cm} \epsfxsize = 6 in
\centerline{\epsfbox{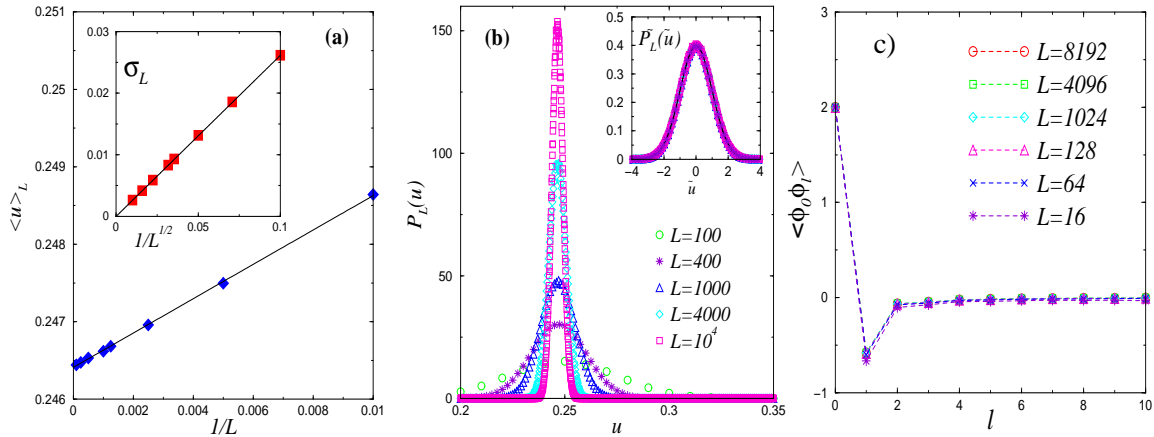}}
\protect\vspace*{-0.2cm}
\caption{a) Steady state average utilization as a function of the number
of PE-s $L$ in a one-dimensional ring geometry; b)
shows that the full distribution
for the rescaled utilization in the steady state
$\tilde{u} = (u(L)-\langle u(L) \rangle)/ \sigma_L$ can be collapsed onto
the normal distribution. c) Slope-slope correlation function.}
\label{Fig:3}
\end{figure}

Both the utilization (density of minima) and the width of the time
horizon are quantities characterizing the fluctuations of the
growing surface. The type of fluctuations can be classified into
universality classes, each class having distinct statistical
properties. Studying the PDES-s scheme as a surface growth model,
we can describe its fluctuations and identify the surface growth
universality class it belongs to. In order to do that we first
introduce the slope variables, $\phi_i = \tau_i - \tau_{i-1}$.
Provided $\tau_i(t)$ is a local minimum, a rod deposited of length
$\eta_i$ corresponds to taking an amount of $\eta_i$ from
$\phi_{i+1}$ and adding it to $\phi_i$, since $\phi_i(t+1) =
\tau_i(t) - \tau_{i-1}(t) + \eta_i(t)$ and $\phi_{i+1}(t+1) =
\tau_{i+1}(t) - \tau_{i}(t) - \eta_i(t)$, i.e., in the {\em
surface of slopes}, $\{\phi_i\}_{i=1}^{L}$ the dynamics is {\em
biased surface diffusion}, given by the equation:
\begin{equation}
\phi_{i}(t+1)-\phi_i(t) = \eta_i(t)\theta(-\phi_i(t))\theta(\phi_{i+1}(t)) -
\eta_{i-1}(t)\theta(-\phi_{i-1}(t))\theta(\phi_{i}(t)) \label{slopes}
\end{equation}
with the constraint $\sum_{i=1}^{L} \phi_i = 0$ generated by the periodic boundary conditions in the $\tau$
variables. In terms of the local slope variables, the
expression for the average density of minima, or average 
utilization becomes: $\langle u(L,t) \rangle = \frac{1}{L} \sum_{i=1}^{L}
\left\langle\theta(-\phi_i(t))\theta(\phi_{i+1}(t))\right\rangle$. Translational invariance implies (no
node is statistically special) $\langle u(L,t) \rangle = \left\langle\theta(-\phi_i(t))
\theta(\phi_{i+1}(t))\right\rangle$ for any $i = 1,2,..,L$. From (\ref{ring}) it follows that
$\langle \tau_{i}(t+1) \rangle - \langle \tau_{i}(t) \rangle = \langle u(L,t) \rangle$, thus, {\em  the average
rate of propagation of the MPEU surface is identical to the average utilization of the PDES-s scheme}. It
is also easy to see that in the slope language it is identical to the average current in the ring.
Next we perform a naive coarse graining by using the representation $\theta(\phi) = \lim_{\kappa \to 0}
\frac{1}{2}\left[ 1+\tanh(\phi/\kappa)\right]$, and keeping only the terms up to the order $(\phi/\kappa)$.
Performing this coarse graining, one obtains:
\begin{equation}
\langle \phi_i(t+1)\rangle - \langle \phi_i(t)\rangle = 
\frac{1}{4\kappa} \langle \phi_{i+1}(t) - 2 \phi_i(t)
+\phi_{i-1}(t)\rangle - \frac{1}{4\kappa^2} \langle \phi_i(\phi_{i+1}-\phi_{i-1}) \rangle\;.
\label{phi_discrete}
\end{equation}
Strictly speaking all of the $(\phi/\kappa)^n$, $n=1,2,...$ terms are
divergent, but taking the proper continuum limit and introducing an
appropriately scaled bias, the only relevant terms can be shown to be those
appearing in Eq.~(\ref{phi_discrete}) \cite{TKXX}. 
In the continuum limit, one thus, obtains for the coarse-grained field:
\begin{equation}
\frac{\partial }{\partial t} \hat{\phi} = \frac{\partial^2 }{\partial x^2} \hat{\phi} - \lambda
\frac{\partial  }{\partial x}\hat{\phi}^2  \label{hatphi}
\end{equation}
where $\lambda$ is a parameter related to the coarse-graining
procedure. The above nonlinear partial differential equation (\ref{hatphi}) is known
as the nonlinear biased diffusion, or the Burgers equation \cite{B74}. 
Returning to the coarse-grained equivalent of the height, or virtual times, $\hat{\tau}$, via
$\hat{\phi} = \partial \hat{\tau}/\partial x$, we obtain the
Kardar-Parisi-Zhang (KPZ) equation \cite{KPZ86}:
%%%%%%%%%%%%%%%%%%%%%%%%%%%%%%%%%%%%%%%%%%%%%%%%%%%%%%%%%%%%%%%%%%%%%%%%%%%%%%%%%%%%%%%%%%%%%
\begin{equation}
\frac{\partial \hat{\tau}}{\partial t}  = \frac{\partial^2 \hat{\tau} }{\partial x^2} - \lambda
\left( \frac{\partial \hat{\tau} }{\partial x} \right)^2\;. \label{hattau}
\end{equation}
%%%%%%%%%%%%%%%%%%%%%%%%%%%%%%%%%%%%%%%%%%%%%%%%%%%%%%%%%%%%%%%%%%%%%%%%%%%%%%%%%%%%%%%%%%%%
To capture the fluctuations, one typically adds a delta-correlated noise term
$\xi(x,t)$, to the right hand side,
conserved for Eq.~(\ref{hatphi}) (i.e., $\int \xi dx = 0$), and non-conserved for Eq.~(\ref{hattau}).
It is important to note that we obtained the KPZ equation as a result of a coarse-graining
procedure. While this results in the ``loss'' of some of the
microscopic details for the original growth model on the lattice,
Eq.~(\ref{hattau}) with noise added captures the long-wavelength
behavior of the MPEU model. Thus, we claim that virtual time horizon
for the basic conservative PDES scheme exhibit kinetic roughening and
it belongs to the KPZ universality class. Identifying the universality
class of a model is one of the main objectives and used extensively in
surface science to classify fluctuation statistics. Our procedure
above indicates that the long-wavelength statistics of the
fluctuations of the time horizon for the basic conservative PDES-s
scheme is in fact captured by the nonlinear KPZ equation. In one
dimension a steady-state for the surface fluctuations is
reached (in the long time limit) for any {\em finite} system-size and
it is governed by the Edwards-Wilkinson (EW) Hamiltonian
${\cal H}_{EW} \propto \int dx \left( \frac{\partial \hat{\tau}}
{\partial x}\right)^2$ (see, e.g., \cite{BS95}). The corresponding surface
is a simple random-walk surface, where the slopes are independent
random variables in the steady state. 
This means that out of the four local configurations of slopes around a point
(down-up, down-down, up-up, up-down) only
one contributes in average to a minimum (down-up),
and since they are all equally likely, we conclude that
$\langle u_{EW}(L\to\infty,t\to\infty) \rangle = 1/4=0.25$.
Our numerical simulations for the MPEU model, see Fig.\ref{Fig:3}a)
indicate a value of $\langle u(L\to\infty,t\to\infty)
\rangle = 0.24641\pm(7\times10^{-6})$ a value close,
but definitively not the same as for the simple random walk surface.
The reason for the obvious difference
is that the coarse-grained version and the original microscopic model are
not identical over the whole spectrum of wavelengths
of the fluctuations. 
%%%%%%%%%%%%%%%%%%%%%%%%%%%%%%%%%%%%%%%%%%%%%%%%%%%%%%%%%%%%%%%%%%%%%%%%%%%
%%%%%%%%%%%%%%%%%%%%%%%%%%%%%%%%%%%%%%%%%%%%%%%%%%%%%%%%%%%%%%%%%%%%%%%%%%%%%
The coarse-graining procedure preserves
the statistics of the long-wavelength modes,
but it looses some information on the short-wavelength ones. In particular the
density of minima is heavily influenced by the
short wavelengths (by how ``fuzzy'' the interface is).
However, the density of minima cannot vanish
in the thermodynamic limit: a zero density of local minima would
imply that it is zero on all length-scales which
would contradict the fact that it belongs to the EW universality
class. The fact that the steady-state of the MPEU model belongs to the EW universality
class guarantees that the local slopes are {\em short-range}
correlated [Fig.3c)], and
that the finite-size corrections for the density of local minima
(average propagation rate of the surface) follows a universal scaling form \cite{KRUG}:
%%%%%%%%%%%%%%%%%%%%%%%%%%%%%%%%%%%%%%%%%%%%%%%%%%%%%%%%%%%%%%%%%%%%%%%%%%
\begin{equation}
\langle u(L,\infty)\rangle \simeq \langle
u(\infty,\infty)\rangle + \frac{\mbox{const.}}{L^{2(1-\alpha)}} \;.
\label{util_scale}
\end{equation}
%%%%%%%%%%%%%%%%%%%%%%%%%%%%%%%%%%%%%%%%%%%%%%%%%%%%%%%%%%%%%%%%%%%%%%%%%%%
Here $\alpha$ is the roughness exponent (equals to $1/2$ for the EW
universality class), characterizing the macroscopic surface-height
fluctuations, as described in detail in the next paragraph.
Figure \ref{Fig:3} confirms this scaling behavior. Further, calculating the spread
in the average utilization the steady state as function of system size,
$\sigma^2_L = \langle u^2(L,\infty)\rangle - \langle u(L,\infty)\rangle^2$, we obtain
$\sigma_L \propto L^{-1/2}$. These findings suggest that the utilization is a
self-averaging macroscopic quantity: its full distribution $P_L(u)$ for large
$L$ is a Gaussian [Fig. \ref{Fig:3}b)].
%%%%%%%%%%%%%%%%%%%%%%%%%%%%%%%%%%%%%%%%%%%%%%%%%%%%%%%%%%%%%%%%%%%%%%%%%%%
\begin{figure}[htbp]
\protect\vspace*{-0.1cm} \epsfxsize = 6 in
\centerline{\epsfbox{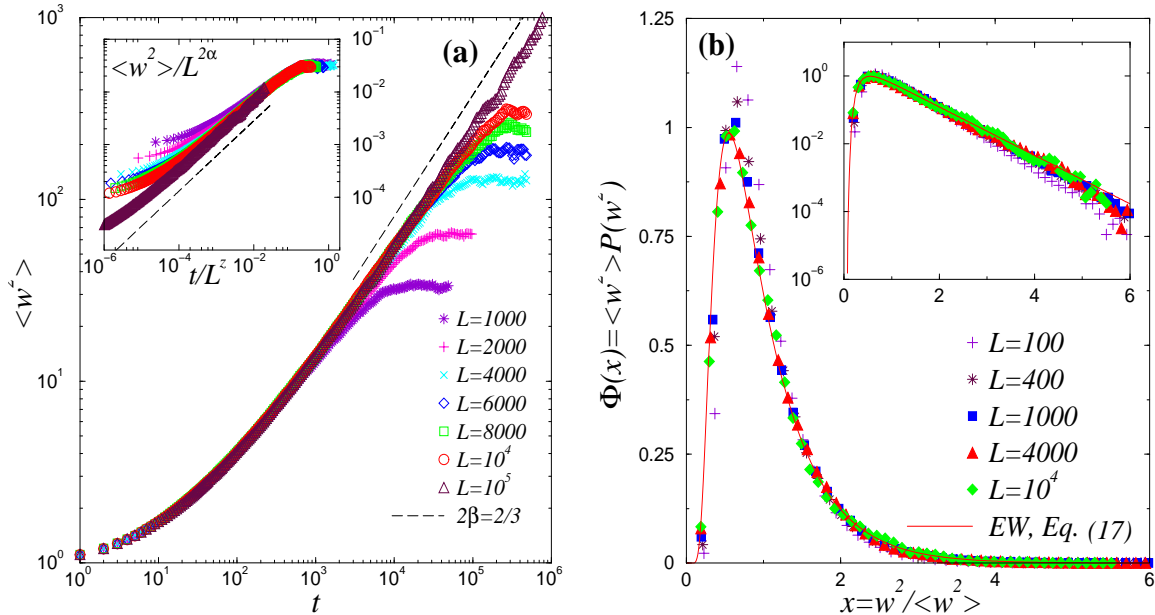}}
\protect\vspace*{-0.2cm}
\caption{a) The width of the time horizon fluctuations show dynamical
scaling and indicate KPZ universality. b) The scaling function for the
steady-state width distribution follows the scaling function
for the EW (one-dimensional KPZ) universality class.}
\label{Fig:4}
\end{figure}
%%%%%%%%%%%%%%%%%%%%%%%%%%%%%%%%%%%%%%%%%%%%%%%%%%%%%%%%%%%%%%%%%%%%%%%%%%

In the following we show numerical results supporting our
claim that the MPEU model belongs to the
KPZ universality class. One of the fundamental characteristic  quantities
strongly influenced by the long-wavelength modes
is the average width of the height fluctuations, as
given in Eq.~(\ref{twidth}). As the surface grows due to
deposition, after an initial transient the growth of
the width will grow as a power law
$\langle w^2(L,t) \rangle \sim t^{2\beta}$ along with
the lateral surface correlations $\xi_{||}(L,t) \sim t^{1/z}$,
until the correlations reach the system size ($\xi_{||} = L$) at a
crossover time $t_{\times}$ (see, e.g., Ref. \cite{BS95}). After the crossover time $t_{\times}$ (for
any finite system $L$) the surface fluctuations are governed by a
steady-state distribution and the width scales as 
%%%%%%%%%%%%%%%%%%%%%%%%%%%%%%%%%%%%%%%%%%%%%%%%%%%%%%%%%%%%%%%%%%%%%%
\begin{equation}
\langle w^2(L,\infty) \rangle \sim L^{2\alpha}\;.
\label{width_scale}
\end{equation}
%%%%%%%%%%%%%%%%%%%%%%%%%%%%%%%%%%%%%%%%%%%%%%%%%%%%%%%%%%%%%%%%%%%%%%
The exponent $\beta$ is called the growth exponent,
$\alpha$ is the called roughness exponent and $z$ is
called the dynamic exponent in the surface growth literature
\cite{BS95}. It is easy to show that the three exponents
are not all independent, and $\alpha = z\beta$ holds \cite{BS95}.
Also, these scaling forms allow one to collapse all the
different curves for the width onto a single function in
the scaling regime, expressing the {\em dynamic scaling
property} of the width: $\langle w^2(L,t) \rangle =
L^{2\alpha} f(t/L^z)$ ($f$ is easy to read of after
comparing it to the scaling behavior). For the KPZ interface,
the analytically obtained exact values for the exponents
are: $\beta=1/3$, $\alpha = 1/2$ and $z = 3/2$.
Fig. \ref{Fig:4} shows the numerically measured scaling
properties for the width of the MPEU model.
the numerically obtained values for the exponents
$\beta = 0.326\pm 0.005$, $\alpha = 0.49 \pm 0.01$ for large
system sizes ($L=10^5$) confirm the KPZ behavior, including
the dynamical scaling property (inset).
Another confirmation for the EW universality class in the
steady state, comes from measuring the full
width distribution $P(w^2)$. For systems belonging to the
EW universality class and having the same type of boundary conditions imposed,
the width distribution has a universal scaling form\cite{FORWZ94}
$P(w^2) = \frac{1}{\langle w^2 \rangle} \Phi\left(\frac{w^2}{\langle
w^2 \rangle}\right)$ 
with
%%%%%%%%%%%%%%%%%%%%%%%%%%%%%%%%%%%%%%%%%%%%%%%%%%%%%%%%%%%%%%%%%%%%%%%%%%%%
\begin{equation}
\Phi(x) = \frac{\pi^2}{3} \sum_{n=1}^{\infty}
(-1)^{n-1} n^2 e^{-\frac{\pi^2}{6}n^2 x}\;, \label{Phi}
\end{equation}
%%%%%%%%%%%%%%%%%%%%%%%%%%%%%%%%%%%%%%%%%%%%%%%%%%%%%%%%%%%%%%%%%%%%%%%%%%%%
where the above scaled distribution is for the case of periodic boundary
conditions.
Figure \ref{Fig:4}b) is a confirmation that MPEU indeed belongs
to the steady state of the EW class, implying that the average
utilization (density of local minima) approaches a non-zero, finite
value in the thermodynamic limit (\ref{uinf}) as reflected by
Eq.~(\ref{util_scale}). Therefore, the basic conservative
scheme is {\em computationally scalable}. 
[For an in-depth and systematic analytical
calculation of the density of minima (utilization)
for a number of surface growth models
(including the EW class) see Ref. \cite{TKDZ00}.]
The measurement phase of the basic conservative
scheme, however, is {\em not} scalable, as indicated by the power-law
divergence of the width in the long-time large $L$ limit, Eq.~(\ref{width_scale}).     
For higher-dimensional topologies, using universality arguments, the conclusion
remains the same: the basic conservative PDES-s is computationally
scalable, but the measurement phase may not be, depending on the upper
critical dimension \cite{BS95} of the surface (see Ref. \cite{KNRGT02,KNTR01}).

\subsection{The conservative scheme on small world networks}

From the previous section it follows that the average
width of the fluctuations scales in the steady state
as $\langle w^2(L,t$$=$$\infty) \rangle \sim L^{2\alpha} = L $ ,
i.e., it grows linearly with the system
size. This means that the basic conservative PDES-s scheme
is {\em not} measurement scalable. Standard methods to control the
width of the virtual time horizon in a PDES scheme utilize some kind
of a windowing technique \cite{FUJI90}. That is, the height of the
local simulated 
time at any PE cannot progress beyond an appropriately chosen and
regularly updated ``cap'', measured from the global
minimum of the time horizon \cite{KKN03}. Thus, a PDES scheme with a
moving window relies on frequent global synchronizations or
communications, which, depending on the architecture, can get costly
for large number of PEs. Here we show how to modify the original
conservative scheme such that the scheme is also measurement scalable
{\em without} global ``intervention'' \cite{KNGTR03}.

The divergence of the width of the surface fluctuations is
closely related to the fact that the lateral surface correlations
also grow with the system size.  In particular, for the
one-dimensional EW surface in the steady state, for large $L$ (and fixed $l$)
%%%%%%%%%%%%%%%%%%%%%%%%%%%%%%%%%%%%%%%%%%%%%%%%%%%%%%%%%%%%%%%%%%%%%%%%%%%%%
\begin{equation}
\langle \hat{\tau}_i \hat{\tau}_{i+l}\rangle \propto  \xi_{||}(L,\infty)-|l|\;,
\end{equation}
%%%%%%%%%%%%%%%%%%%%%%%%%%%%%%%%%%%%%%%%%%%%%%%%%%%%%%%%%%%%%%%%%%%%%%%%%%%%%%
where $\hat{\tau}_i$ are the coarse-grained height fluctuations
measured from the mean and $\xi_{||}(L,\infty)\sim L$. Thus, 
$\langle w^2(L,\infty)\rangle =\langle\hat{\tau}_i^2\rangle\propto\xi_{||}(L,\infty)
\sim L$.
The ``height-height'' correlations can be characterized by introducing
the structure factor for the heights: 
%%%%%%%%%%%%%%%%%%%%%%%%%%%%%%%%%%%%%%%%%%%%%%%%%%%%%%%%%%%%%%%%%%%%%%%%%%%%%%
\begin{equation}
S^{(\tau)}(k) = \frac{1}{L} \langle
\tilde{\tau}_k \tilde{\tau}_{-k}\rangle \label{structf}
\end{equation}
%%%%%%%%%%%%%%%%%%%%%%%%%%%%%%%%%%%%%%%%%%%%%%%%%%%%%%%%%%%%%%%%%%%%%%%%%%%%%%
where $k=2\pi n/L$, $n=0,1,2,..,L-1$ is the wave-vector, and
$\tilde{\tau}_k = \sum_{j=0}^{L-1} e^{-ikj} (\tau_j-\overline{\tau})$ is the discrete spatial
Fourier transform of the fluctuations of virtual time horizon. Then
%%%%%%%%%%%%%%%%%%%%%%%%%%%%%%%%%%%%%%%%%%%%%%%%%%%%%%%%%%%%%%%%%%%%%%%%%%%%%%%
\begin{equation}
\langle \hat{\tau}_i \hat{\tau}_{i+l}\rangle =
\frac{1}{L} \sum_{k} e^{ikl} S^{(\tau)}(k)\; 
\label{Sk}
\end{equation}
%%%%%%%%%%%%%%%%%%%%%%%%%%%%%%%%%%%%%%%%%%%%%%%%%%%%%%%%%%%%%%%%%%%%%%%%%%%%%
and 
%%%%%%%%%%%%%%%%%%%%%%%%%%%%%%%%%%%%%%%%%%%%%%%%%%%%%%%%%%%%%%%%%%%%%%%%%%%%%
\begin{equation}
\langle w^2(L,\infty)\rangle  = \frac{1}{L} \sum_{k} S^{(\tau)}(k)\;. 
\label{wSk}
\end{equation}
%%%%%%%%%%%%%%%%%%%%%%%%%%%%%%%%%%%%%%%%%%%%%%%%%%%%%%%%%%%%%%%%%%%%%%%%%%%%
Since the universality class for the time horizon evolution is EW, it
follows that the expected behavior for the steady-state structure
factor for small wave-numbers is 
%%%%%%%%%%%%%%%%%%%%%%%%%%%%%%%%%%%%%%%%%%%%%%%%%%%%%%%%%%%%%%%%%%%%%%%%%%%%%
\begin{equation}
S^{(\tau)}(k) \propto \frac{1}{k^2}\;
\end{equation}
%%%%%%%%%%%%%%%%%%%%%%%%%%%%%%%%%%%%%%%%%%%%%%%%%%%%%%%%%%%%%%%%%%%%%%%%%%%%
(see, e.g., Eq.~(11) in
Ref. \cite{TKDZ00}). 
%%%%%%%%%%%%%%%%%%%%%%%%%%%%%%%%%%%%%%%%%%%%%%%%%%%%%%%%%%%%%%%%%%%%%%%%%%%
\begin{figure}[htbp]
\protect\vspace*{-0.1cm} \epsfxsize = 4 in
\centerline{\epsfbox{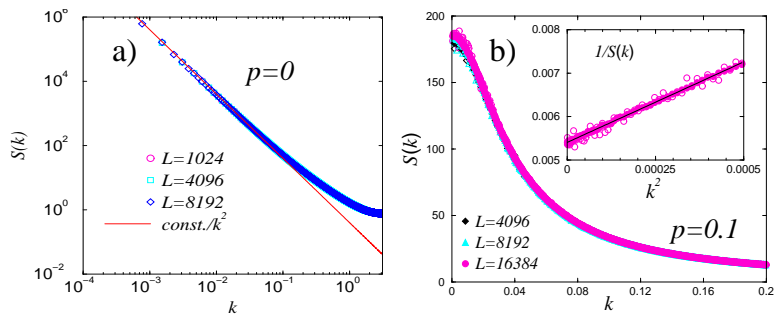}}
\protect\vspace*{-0.2cm}
\caption{Steady-state structure factors for the virtual time horizon for
the a) basic conservative scheme on a regular one-dimensional lattice
($p$$=$$0$) and b) for the small-world scheme with $p$$=$$0.1$.}
\label{Fig:5}
\end{figure}
%%%%%%%%%%%%%%%%%%%%%%%%%%%%%%%%%%%%%%%%%%%%%%%%%%%%%%%%%%%%%%%%%%%%%%%%%%%
Indeed, this is also confirmed by our direct simulation results, shown
Fig.~\ref{Fig:5}a). 
This form of the structure factor implies that there are no
lengthscales other than the lattice constant and the systems size, and
thus, the correlation length and the width diverge in the
thermodynamic limit, as also can be seen by directly evaluating
Eq.~(\ref{wSk}). 
%%%%%%%%%%%%%%%%%%%%%%%%%%%%%%%%%%%%%%%%%%%%%%%%%%%%%%%%%%%%%%%%%%%%%%%%%%%%%%%
\begin{figure}[htbp]
\protect\vspace*{-0.1cm} \epsfxsize = 4 in
\centerline{\epsfbox{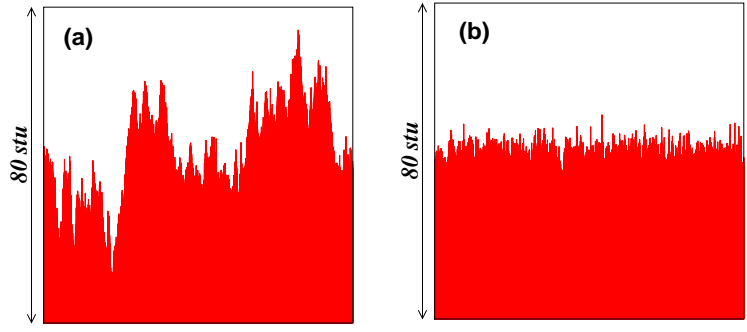}}
\protect\vspace*{-0.2cm}
\caption{Steady-state  virtual time horizon snapshots
with $L=10,000$ after $t=10^6$ parallel algorithmic steps (Monte-Carlo sweeps)
for a) the basic conservative scheme ($p=0$) and b) the small-world scheme
$p=0.1$. Note that the vertical scales are the same in a) and b)
(plotted in arbitrary simulated time units [stu]).}
\label{Fig:6}
\end{figure}
%%%%%%%%%%%%%%%%%%%%%%%%%%%%%%%%%%%%%%%%%%%%%%%%%%%%%%%%%%%%%%%%%%%%%%%%%%%%%%

To de-correlate the surface fluctuations, we modify the communication
topology in the following way \cite{KNGTR03}:
for every node $i$, at the onset of the simulation, we introduce
one extra quenched random communication link $r(i)$.
Together with the existing regular topology, these extra
communication links will form a small-world
graph \cite{WS98,NW99,K00}. Note that in our specific construction of
the small-world network, each node has exactly one random connection and
$r(r(i))$$=$$i$, so that there are exactly $L/2$ random links distributed.
The updating on PE $i$ will obey the following probabilistically
chosen condition:
\begin{eqnarray}
 \tau_i \leq
\left\{
 \begin{array}{l}
   \min\{\tau_{i-1},\tau_{i+1},\tau_{r(i)}\}\;\;\;\mbox{with probability}\;\;p \\
   \min\{\tau_{i-1},\tau_{i+1}\}\;\;\;\mbox{with probability}\;\;1-p
 \end{array}
 \right. \;.
\label{csw}
\end{eqnarray}
The PE {\em actually performs the update} (generate the virtual time of
the next update, or deposit the rod at $i$
in the MPEU surface) if condition (\ref{csw}) is fulfilled.
This means that for sites that would normally be updated
within the basic conservative scheme, i.e., $ \tau_i \leq
\min\{\tau_{i-1},\tau_{i+1}\}$ the PE will make an
extra check for the condition $ \tau_i \leq \tau_{r(i)}$ with probability $p$.
The parameter $p$  allows us to tune the scalability properties of the
corresponding PDES scheme on the quenched small-world 
network continuously from the pure basic conservative scheme ($p=0$)
to the ``fully'' small-world conservative scheme ($p=1$). 
%%%%%%%%%%%%%%%%%%%%%%%%%%%%%%%%%%%%%%%%%%%%%%%%%%%%%%%%%%%%%%%%%%%%%%%%%%
\begin{figure}[htbp]
\protect\vspace*{-0.1cm} \epsfxsize = 5 in
\centerline{\epsfbox{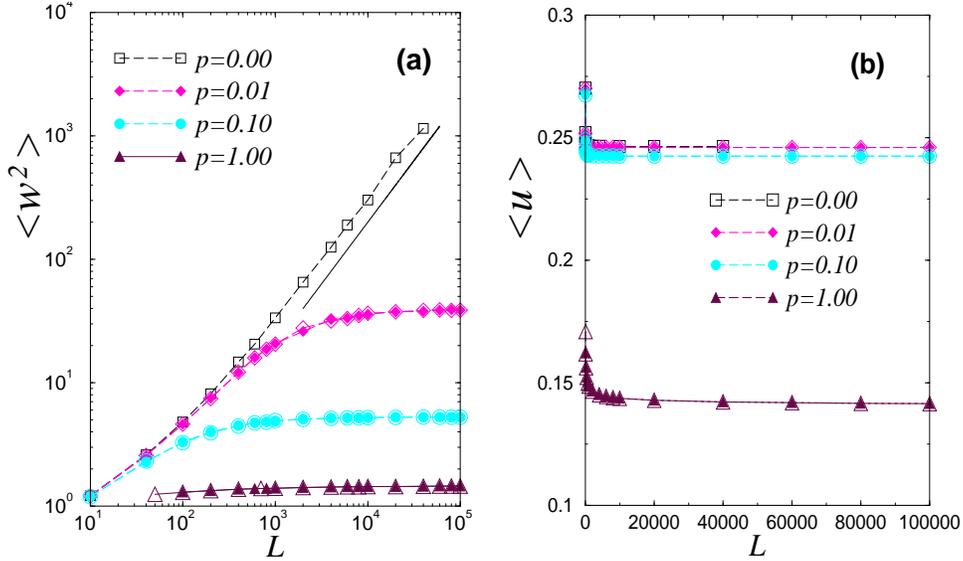}}
\protect\vspace*{-0.2cm}
\caption{a) The average steady-state width and b) the utilization for various
$p$ values. In addition to ensemble averages over 10 realizations
of the random links (solid symbols) a single realiztion is also shown
(open symbols). The solid straight line has a slope of $1/2$ and represents
the asymptotic one-dimensional KPZ power-law divergence of the
width for the basic conservative scheme ($p=0$).}
\label{Fig:7}
\end{figure}
%%%%%%%%%%%%%%%%%%%%%%%%%%%%%%%%%%%%%%%%%%%%%%%%%%%%%%%%%%%%%%%%%%%%%%%%5
These occasional extra checkings through the quenched random links
are not necessary for the faithfulness of the simulation. It is merely
used to {\em synchronize} the PEs in such a way that the fluctuations of the
time horizon remain bounded in the infinite system-size limit.
Most importantly, as the width is reduced from ``infinity'' (or some
large number proportional to $L$ for finite number of PEs) to a finite
controlled value, the utilization still remains bounded away from zero.

To support this statement, we first use the same
coarse-graining procedure used to derive the KPZ
equations as the continuum counterpart of the
MPEU model. For the small-world topology we obtain
%%%%%%%%%%%%%%%%%%%%%%%%%%%%%%%%%%%%%%%%%%%%%%%%%%%%%%%%%%%%%%%%%%%%%%%%%%%%%
\begin{equation}
\frac{\partial \hat{\tau}}{\partial t}  =
-\gamma(p)\hat{\tau} + \frac{\partial^2 \hat{\tau} }{\partial x^2} - \lambda
\left( \frac{\partial \hat{\tau} }{\partial x} \right)^2 + \xi(x,t)\;. 
\label{hattausw}
\end{equation}
%%%%%%%%%%%%%%%%%%%%%%%%%%%%%%%%%%%%%%%%%%%%%%%%%%%%%%%%%%%%%%%%%%%%%%%%%%%%%
with $\gamma(p) = 0$ for $p=0$, and $\gamma(p) > 0$ for $0 < p \leq 1$.
This implies that the extra checking along the random links introduces a {\em strong}
relaxation (first term on the rhs of (\ref{hattausw})) for the
long-wavelength modes of the surface fluctuations, resulting in a
finite width. A more transparent picture is gained if we look at the
steady-state structure factor (\ref{structf}). Restricting our
attention to the linear terms in Eq.~(\ref{hattausw}) we obtain
%%%%%%%%%%%%%%%%%%%%%%%%%%%%%%%%%%%%%%%%%%%%%%%%%%%%%%%%%%%%%%%%%%%%%%%%%%%%%%%%
\begin{equation}
S^{(\tau)}(k) \propto \frac{1}{\gamma+k^2} 
\label{sfsw}\;.
\end{equation}
%%%%%%%%%%%%%%%%%%%%%%%%%%%%%%%%%%%%%%%%%%%%%%%%%%%%%%%%%%%%%%%%%%%%%%%%%%%%%%%
In this approximation, the lateral correlation length $\xi_{||}$ scales as
$1/\sqrt{\gamma}$, and remains {\em finite} (and system-size
independent) in the thermodynamic limit for all $p>0$ (i.e., for an 
arbitrary small probability to utilize the random links). 
Figure \ref{Fig:5}b) shows the structure factor for the small-world network
with $p=0.1$, confirming the prediction of Eq.~(\ref{sfsw}) for small
wave numbers. Consequently, the height-height correlations decay
exponentially 
%%%%%%%%%%%%%%%%%%%%%%%%%%%%%%%%%%%%%%%%%%%%%%%%%%%%%%%%%%%%%%%%%%%%%%%%%%%%%
\begin{equation}
\langle \hat{\tau}_i \hat{\tau}_{i+l}\rangle \propto \xi_{||}\; e^{-|l|/\xi_{||}}\;,
\end{equation}
%%%%%%%%%%%%%%%%%%%%%%%%%%%%%%%%%%%%%%%%%%%%%%%%%%%%%%%%%%%%%%%%%%%%%%%%%%%%%%
and the width remains finite, 
$\langle w^2(L,\infty)\rangle \sim\xi_{||}$, where $\xi_{||}$ is
independent of the systems size for all $p>0$. Further, for the structure factors of
the local slopes (the Fourier transform of the slope-slope
correlations) one obtains 
%%%%%%%%%%%%%%%%%%%%%%%%%%%%%%%%%%%%%%%%%%%%%%%%%%%%%%%%%%
\begin{equation}
S^{(\phi)}(k) = \frac{1}{L} \langle\tilde{\phi}_k
\tilde{\phi}_{-k}\rangle =
k^2 S^{(\tau)}(k) \propto 
\frac{k^2}{\gamma+k^2} = 1 - \frac{\gamma}{\gamma+k^2} \;.
\end{equation}
%%%%%%%%%%%%%%%%%%%%%%%%%%%%%%%%%%%%%%%%%%%%%%%%%%%%%%%%%%%
Both terms above yield short-range correlations (delta function for
the first term and exponential decay for the second one), thus, the
slopes remain short-range correlated, resulting in a non-zero density
of local minima.
Figure \ref{Fig:6} shows two snapshots of
the virtual time horizons for the basic conservative scheme $p=0$, and the
small-world scheme with $p=0.1$.
Figure \ref{Fig:7}a) shows the scaling of the steady state width
with the system size
for various $p$ values and Fig.\ref{Fig:7}b)
shows the scaling of the average, steady
state utilization with the system
size for the same set of $p$ values.
Notice that when increasing $p$ (from $p$$=$$0$ to $p$$=$$0.01$), the width
instantaneously drops from a linear divergence to a saturated value,
while at the same time,  the utilization hardly changes.
In fact, an infinitesimally
small $p$ will make the width bounded, but
at an only infinitesimal expense to the utilization. For example, for
a hypothetically infinite system, taking
$p$$=$$0.01$, the width is reduced from infinity to about $40$, while
the utilization from $0.2464$ only to about $0.246$;
for $p$$=$$0.1$, the width is further reduced to
about $5$, while the utilization only to $0.242$.
By further increasing $p$, the width further reduces,
and at $p$$=$$1$ it is  about $1.46$, whereas the
utilization decreases to $0.141$, still clearly bounded away from zero
in the thermodynamic limit.

\section{Scalability of the Conservative PDES-s scheme
on scale-free network topologies} \label{sec3}

The Internet is a spontaneously grown collection
of connected computers. The number of (only) webservers
by February 2003 reached
over 35 million \cite{netcraft}. The number of PC-s in use
(Internet users) surpassed 660 million in 2002, and it is
projected to surpass one billion by 2007 \cite{cia}.
\begin{figure}[htbp]
\protect\vspace*{-0.1cm} \epsfxsize = 3.0 in
\centerline{\epsfbox{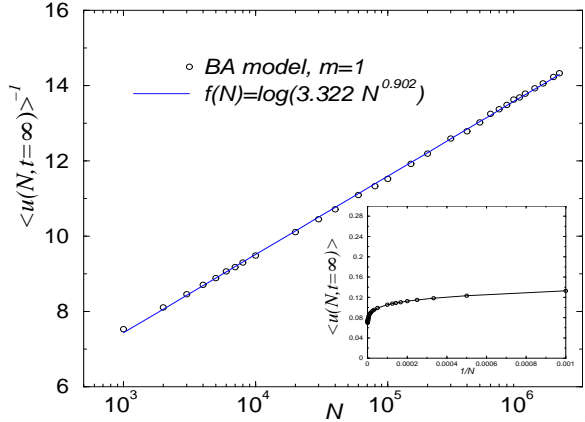}}
\protect\vspace*{-0.2cm}
\caption{Steady-state utilization for the scale-free BA model.}
\label{uinfg}
\end{figure}
The idea for using it as a giant supercomputer is
rather natural: many computers are in an idle state, running
at best some kind of screen-saver software, and
the ``waisted" computational time is simply immense. Projects such
as SETI@home \cite{SETI} or the GRID consortium
\cite{GRID} are targeting to harness the power lost in screen-savers.
\begin{figure}[htbp]
\protect\vspace*{-0.1cm} \epsfxsize = 3.4 in
\centerline{\epsfbox{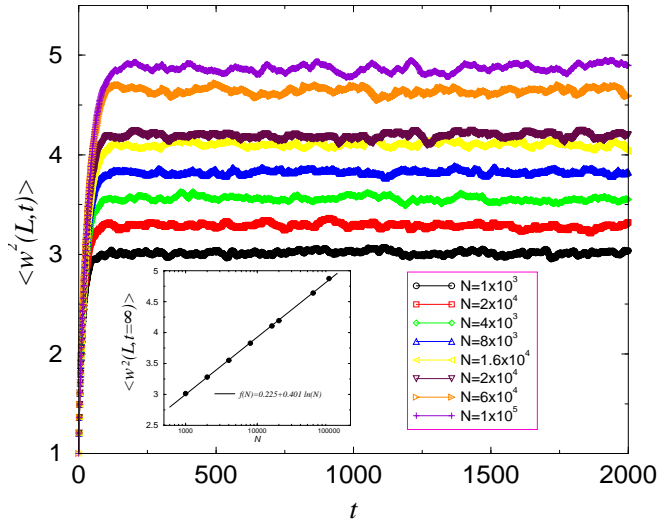}}
\protect\vspace*{-0.2cm}
\caption{Behavior of the time horizon width for the scale-free BA network.
The inset shows the scaling of the steady-state width as function
of system size, $N$.
Notice the log-lin scale on the inset.}
\label{width1}
\end{figure}

Most of the problems solved currently with
distributed computation on the Internet are ``embarrassingly parallel"
\cite{K03} , i.e., the computed tasks have
little or no connection to each other similar to starting the
same run with a number of different random seeds,
and at the end collecting the data to perform
statistical averages.  However, before more
large-scale, complex problems can be solved in real time on the Internet
a number of challenges have to be solved, such as
the task allocation problem which is rather complex by itself
\cite{S99}.

Here we ask the following question: {\em Given}
that task allocation is resolved and the PE communication
topology on the internet is a scale-free network,
what are the scalability properties of a PDES-s scheme on
such networks? Here we present numerical results,
for the PDES-s update scheme, as measured on a model of
scale-free networks, namely the Barab\'asi-Albert
model \cite{BA99,BAJ99}. This network is created through the
stochastic process of preferential attachment: to
the existing network at time $t$ of $N$ nodes,  attaches
the $N+1$-th node with $m$ links (``stubs") at time
$t+1$, such that each stub attaches to a node with probability
proportional to the existing  degree (at $t$) of the
node. Here we will only present the $m=1$ case, when the
network is a scale-free tree. Once we reach a given number of nodes
in the network, we stop the process and use the random network instance
to run the MPEU model on  top of it, using the evolution equation
(\ref{eveq}) for the time horizon.
 While in case of regular topologies, the degree of a node is constant,
e.g. for $d$-dimensional ``square'' lattices,
$P^{(L^d)}(k)=2d\delta_{k,2d}$ , for the
BA network, it is a power law in the asymptotic ($N\to \infty$) limit :
$P^{BA}(k)\simeq 2m^2 k^{-3}$. 
%%%%%%%%%%%%%%%%%%%%%%%%%%%%%%%%%%%%%%%%%%%%%%%%%%%%%%%%%%%%%%%%%%%%%%%%%%%%
\begin{figure}[htbp]
\protect\vspace*{-0.1cm} \epsfxsize = 4.5 in
\centerline{\epsfbox{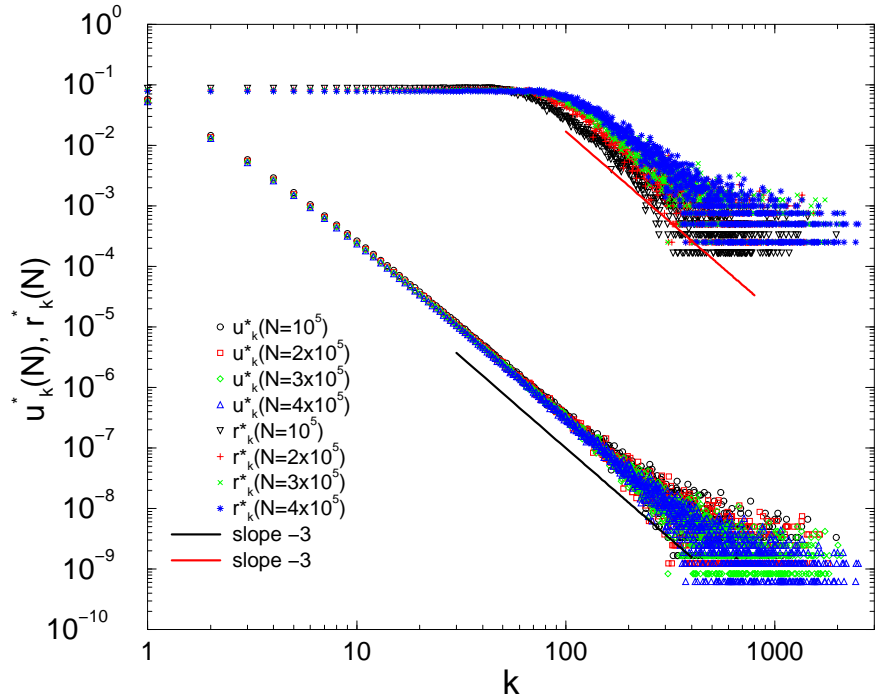}}
\protect\vspace*{-0.2cm}
\caption{Connectivity-utilization $u_k$ and relative connectivity utilization $r_k$
as function of degree. Each data set is obtained after averaging over 200 independent
runs.} \label{ukrk}
\end{figure}
%%%%%%%%%%%%%%%%%%%%%%%%%%%%%%%%%%%%%%%%%%%%%%%%%%%%%%%%%%%%%%%%%%%%%%%%%%%
The condition (\ref{active}) for a site to be
updated, i.e., that its virtual time is a local minimum, is a {\em
local} property, and thus we expect that the utilization itself be
correlated with local structural properties of the graph, such as the
degree distribution. To get a more detailed picture, we define two
more quantities:\\ 
1) {\em connectivity-utilization}:
\begin{equation}
u_k(N,t)= \frac{|{\cal A}_k(t)|}{N} \label{uk}
\end{equation}
which is the fraction of active nodes of degree $k$,
and\\
2) the {\em relative connectivity-utilization}:
\begin{equation}
r_k(N,t)= \frac{|{\cal A}_k(t)|}{N_k} \label{rk}
\end{equation}
which is the fraction of active nodes of degree $k$ within the set of all nodes
of degree $k$. From the abobe definitions we find the following ovious
relations:
$\sum_{k}u_k(N,t)=u(N,t)$ and $\sum_{k} r_k(N,t) N_k / N =
\sum_{k} r_k(N,t) P^{BA}(k) = \sum_{k} u_k(N,t) = u(N,t) = 
\langle r_k(N,t) \rangle_{network} $ at all times.
Figure \ref{uinfg} shows the steady state ($t\to \infty$, in the MPEU model
on a fixed BA network of $N$ nodes) values of the average utilization
as function of the network size $N$. The inset in Fig.~\ref{uinfg} is
analogous to Fig.~\ref{Fig:3}(a) which showed the same quantity on a
ring. Notice that strictly speaking, the PDES-s scheme is
computationally {\em non-scalable}. An empirical
fit suggest that
$u^*(N)=\langle u(N,t$$=$$\infty)\rangle \simeq
\left[\ln{\left(a N^{b}\right)}\right]^{-1}$ 
with $a\simeq 3.322$ and $b=0.902$,  
i.e., the computation is only logarithmically (or
marginally) non-scalable. For a system of $N$$=$$10^3$ nodes we have found
a steady state utilization (for the worst case scenario) of 
$u^*(10^3)=0.1328$ ($13.3\%$ efficiency), 
while for a system of a million nodes, $N$$=$$10^6$, the utilization
dropped only to $u^*(10^6) =0.073$ ($7.3\%$ efficiency), by less than half of its value!
For practical purposes the PDES-s scheme
can be considered computationally scalable, and this
type of non-scalability we will call {\em logarithmic} (or marginal)
non-scalability.

Figure \ref{width1} shows the scaling of the width of the fluctuations
for the time horizon as function of time, and the scaling of its value
in the steady-state as function of system size (inset). Notice, that
while the steady state width diverges to infinity, it only does so
logarithmically, $\langle w^2(N,t$$=$$\infty) \rangle 
\simeq \left[\ln{\left(cN^{d}\right)}\right]$ 
with $c\simeq1.25$ and $d=\simeq 0.401$. 
Some specific values: $\langle w^2(10^3,t$$=$$\infty) \rangle\simeq 3.01$,
$\langle w^2(10^5,t$$=$$\infty) \rangle\simeq 4.78$. This means that
the measurement phase of the PDES-s scheme on a scale-free network is
non-scalable either, however, it is so only logarithmically, and for
practical purposes the scheme can be considered scalable. 
Overall, the PDES-s update scheme has logarithmic (or marginal) non-scalability
on scale-free networks. If one examines the connectivity-utilization
and relative connectivity-utilization in the steady state, as shown in
Fig. \ref{ukrk}, one finds that with good approximation
$u^*_k(N)\sim k^{-3}$, and $r^*_k(N) = const. \simeq
u^*(N)$ for $k \leq k_{\times}$ and
$r^*_k(N)\sim k^{-3}$ for $k > k_{\times}$, with
$k_{\times}\sim 1/u^*(N) = \ln{\left(a N^{b}\right)} \sim \ln{N}$, being a crossover
degree.

\section{Conclusions}
\label{sec4}

We studied the fundamental scalability problem of conservative PDES schemes
where events are self-initiated and have identical distribution on each PE. 
First, we considered the scalability of the basic conservative scheme
for systems with short-range interactions on regular lattices. By
exploiting a mapping \cite{KTNR00} between the progress of the simulation and
kinetic roughening in non-equilibrium surfaces, we found that while
the average progress rate of the PEs  
$\langle u(\infty,\infty)\rangle$ is a finite {\em non-zero}
value, the spread of the progress of the PEs about the mean 
$\langle w^2(\infty,\infty)\rangle$ {\em diverges}. 
The former property makes the measurement phase of the
algorithm non-scalable. 
In order to make the measurement part of the simulation scalable as well, we
introduced \cite{KNGTR03} quenched random connections between PEs (exactly
one for each) so that the resulting random links on the top of the
regular short-range connections formed a {\em small-world}-like connection
topology. Invoking the same conservative protocol used at an
arbitrarily small rate through the random links was sufficient to achieve full
scalability: the PEs progress at a non-zero, near uniform rate {\em without}
requiring global synchronization \cite{KNGTR03}.  
The above construction of a fully scalable algorithm for simulating
large systems with asynchronous dynamics and short-range interactions
is an example for the enormous ``computational power and
synchronizability'' \cite{WS98} that can be achieved by small-world
couplings.
The suppression of critical fluctuations of the virtual time horizon is also 
closely related to the emergence of mean-field-like phase transitions and
phase ordering in {\em non-frustrated} interacting systems 
\cite{BARRAT,GITTERMAN,XY_sw,ising_sw,phase_sw}. 
In particular, the fluctuations exhibited by the virtual time horizon
with small-world synchronization should exhibit very similar
characteristics to the fluctuations of the order parameter in the
XY-model placed on a small-world network \cite{XY_sw}. 

Second, we studied the scalability properties for a causally
constrained PDES scheme hosted by a network of computers where the
network is {\em scale-free} following a ``preferential attachment''
construction \cite{BA99,BAJ99}. Here the PEs simply have to satisfy the general
criterion Eq.~(\ref{active}) in order to advance their local time. 
Despite some nodes in the network having abnormally large connectivity
(as a result of the scale-free nature of the degree distribution), we
found that the computational phase of the algorithm is only
marginally non-scalable. The utilization exhibited slow
logarithmic decay as a function of the number of PEs. At the same
time, the width of the time horizon diverged logarithmically slowly,
rendering the measurement phase of the simulations marginally
non-scalable as well. An intriguing question to pursue is how the
logarithmic divergence of the surface fluctuations observed here can be
related to the collective behavior (in particular, the finite-size
effects of the magnetic susceptibility) of Ising ferromagnets on
scale-free networks \cite{Stauffer,LVVZ,DGM,Bianconi} with the same
degree distribution.

\section*{Acknowledgements}
Discussions with P.A. Rikvold, B.D. Lubachevsky, Z. R\'acz and G. Istrate are
gratefully acknowledged. Z.T. was supported by the DOE under contract
W-7405-ENG-36.
This research is supported in part by NSF
through Grant No.\ DMR-0113049 and the Research Corporation through
Grant No.\ RI0761.

\end{document}